\documentstyle[natbib,aas2pp4]{article}
\input{psfig}

\slugcomment{To appear in Physics Letters} \lefthead{I. Lopes and
J. Silk} \righthead{The WIMPs model constrained by the solar
standard model}
\date{\today}

\begin{document}

\title{\bf SOLAR NEUTRINOS: PROBING THE
QUASI-ISOTHERMAL SOLAR CORE PRODUCED BY SUSY DARK MATTER PARTICLES}

\author{IL\'IDIO P. LOPES\altaffilmark{1,2}, JOSEPH SILK\altaffilmark{1}}

\affil{Department of Physics, Nuclear and Astrophysics
                Laboratory, Keble Road, Oxford OX1 3RH, United Kingdom}

\altaffiltext{1}{Inquiries can be sent to {\bf
lopes@astro.ox.ac.uk} and {\bf silk@astro.ox.ac.uk}}
\altaffiltext{2}{Instituto Superior T\'ecnico, Centro
Multidisciplinar de Astrof\'\i sica,
 Av. Rovisco Pais, 1049-001 Lisboa, Portugal}

\setcounter{page}{1}
\newcommand{\RON}{read-out noise}
\newcommand{\beqa}{\begin{eqnarray}}
\newcommand{\eeqa}{\end{eqnarray}}
\newcommand{\tnew}[1]{{\bf #1}}

%
\begin{abstract}
SNO measurements strongly constrain the central temperature of the
Sun, to within a precision of  much less than 1\%. This result can
be used to constrain the parameter space of SUSY dark matter
particle candidates. In this first analysis we find a lower limit
for the WIMP mass of 60 GeV, well above  the WIMP evaporation
limit of 10 GeV. Furthermore, in the event that WIMPs create a
quasi-isothermal core within the Sun, they will produce a peculiar
distribution of  the solar neutrino fluxes measured on Earth.
Typically, a WIMP with a mass of 100 GeV and annihilation
cross-section of  $10^{-34}\;cm^3/sec$,  will decrease the
standard solar model neutrino predictions, by up to $4\%$ for the
Cl, by $3\%$ for the heavy water, and by  $1\%$ for the Ga
detectors.
\end{abstract}
\keywords{Key~words: stars: oscillations - stars: interiors - Sun:
oscillations - Sun: interior: cosmology - dark matter}
\date{\today}
\twocolumn

In the last three decades, solar neutrino experiments  have
measured fewer neutrinos than were predicted by the solar models.
One explanation for the deficit is the transformation of the Sun's
electron-type neutrinos into other active flavors (Bahcall 1989).
Recently, the Sudbury Neutrino Observatory (SNO) has measured the
$^8B$ solar neutrinos. The results obtained establish direct
evidence for the non-electron flavor component in the solar
neutrino flux and yield the first unequivocal determination of the
total flux of $^8B$ neutrinos produced by the Sun (Ahmad {\it et
al.} 2001; Bahcall 2001).
 The total flux of active $^8B$ neutrinos is
determined to be $5.44\pm 0.99\times 10^{6}\;cm^{-2}s^{-1}$, only
$10\%$ above the theoretical prediction (Turck-Chi\`eze {\it et
al.} 2001) and  consequently in excellent agreement with the
predictions of the different solar standard models (Bahcall,
Pinsonneault and Basu 2001; Turck-Chi\`eze, Nghiem, Couvidat \&
Turcotte 2001). Furthermore, this result is also consistent with
the measurements in the Super-Kamiokande detector (Fukuda {\it et
al.} 2001).

This work discusses the possibility of using the present $^8B$
neutrinos produced in the Sun to determine a lower limit for the
mass of WIMPs,  under the standard hypothesis that the Sun's
evolution takes place within a halo of non-baryonic dark matter in
the form of weakly interacting particles massive (WIMPs).
Furthermore, we predict the corresponding neutrino fluxes that
should be measured on Earth if the presence of dark matter inside
the Sun modifies its evolution.


\begin{figure}[t]
\centerline{
\psfig{file=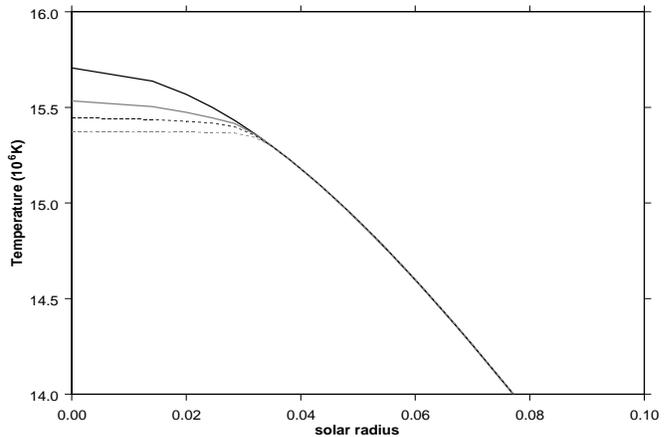,width=10.0cm,height=7.0cm}}
\caption{\small This figure presents the  variation of the radial
distribution of the temperature on the Sun's nuclear region, for
the solar standard model and solar models with different
concentration of WIMPs in the solar core. The standard evolution
of the Sun is represented by the black continuous curve. In the
WIMP-accreting models,  the evolution of the Sun occurs within an
halo of WIMPs with a mass of 100 GeV and annihilation
cross-section of $10^{-34}\;cm^3/sec$. The different curves
correspond to the following scalar scattering cross-sections:
$10^{-38} cm^2$ (grey continuous curve), $10^{-36}cm^2$ (black
dashed curve) and $10^{-40}\;cm^2$ (grey dashed curve). }
\end{figure}

The dynamic behavior of various astronomical objects, from
galaxies to galaxy clusters and to large-scale structures in the
observed universe, can only be understood if the dominant
component of the mean matter density is dark. The bulk of the dark
matter is believed to be non-baryonic, and the existence of
particles that interact with ordinary matter on the scale of the
weak force, WIMPs, arising from the lightest stable particle
predicted by SUSY, provides one of the best candidates to solve
this problem (Jungman,  Kamionkowski \& Griest 1996). In the
current cosmological scenario, we are interested in understanding
the evolution of the Sun within a halo of non-baryonic dark
matter. We assume that the star is in hydrostatic equilibrium, is
spherically symmetric and that the effects of rotation and of
magnetic fields can be neglected. The present structure of the Sun
is obtained by evolving an initial star  within a halo of WIMPs,
from the pre-main sequence,  0.05 Gyr before the ZAMS, until its
present age, 4.6 Gyr (Lopes, Silk \& Hansen 2001). In such
conditions, the WIMPs with masses above 10 GeV accumulate in the
center of the star due to their capture by the Sun's gravitational
field. Consequently, the present abundance of WIMPs inside the
star depends on the number of WIMPs accumulated in the Sun by
capture from the Galactic halo and depleted by annihilation. The
WIMPs captured during the Sun's evolution are confined in the
central part of the nuclear-reacting core of the Sun. The WIMP
core radius is inversely proportional to the square root of the
mass of the WIMPs (Press and Spergel 1985). WIMP accumulation in
the stellar center provides an additional mechanism for
transferring radiative energy from the solar core, changing its
structure locally. In the particular case of WIMPs with very small
scattering cross-sections, or very large mean free paths, the
evacuation of the energy of the Sun's core becomes extremely
efficient, and the core becomes almost isothermal (cf. Fig.1).
Fig.2a presents the variation of the central temperature of the
present Sun in the case of solar models, including WIMPs, relative
to the solar standard model. As expected, the maximum effect
occurs for stellar models with a scattering cross-section of the
order of the critical cross-section, $m_p\;R_\star^2/M_\star$,
where $m_p$ is the proton mass, $R_\star$ is the total stellar
radius and $M_\star$ is the total stellar mass. In the case of the
Sun, the critical cross-section is of the order of
$10^{-36}\;cm^2$, corresponding to a mean free path of the order
of the solar radius. Furthermore, we notice that for very large
annihilation cross-sections, the concentration of WIMPs in the
core of the star is very small, the structure of the Sun is
similar to the solar standard model and, consequently, the central
temperature does not change.

 The neutrino and energy production in the Sun takes place within
$30 \%$ of the solar radius. The strong dependence of the nuclear
reaction rates on the temperature allows us to use  neutrino
nuclear reactions such as $pp$, $^7Be$ and $^8B$ that occur
 at various locations in the nuclear region to infer the
radial distribution of the temperature in the core.

The production of $^8B$ takes place in the inner $2\%$ of the mass
of the solar  core.  $^8B$ decay reactions present the strongest
dependence on the temperature: $^8B$ neutrino production is
maximum at quite small radii, $5\%$
 of the solar radius, and its
generation is confined to  the region between $2\%$ and $7\% $ of
the solar radius.  Consequently, this flux of neutrinos becomes
the best probe  of the temperature at the center of the Sun. Since
the temperature dependence of the $^8B$ neutrino flux is strong,
the total flux of $^8B$ neutrino production can be expressed as a
function of the central temperature of the Sun. Bahcall and Ulrich
(1988) computed the central temperature dependence of the $^8B$
neutrinos for $10^3$ solar models, and found $\phi(^8B)\propto
T_c^{18}$. If the SNO measurement of $^8B$ neutrino fluxes is
correct, the central temperature of the standard solar model is
within  less than $0.5\%$ of the temperature deduced from the
measured $^8B$ neutrino flux. The central temperature of the Sun
is therefore estimated to be approximately $15.78\;\;10^6\;K$.
This result is confirmed by the  computation of
 Fiorentini, Villante and Ricci (2001),
using the combined results of SNO and Super-Kamiokande.

\begin{figure}
\centerline{\psfig{file=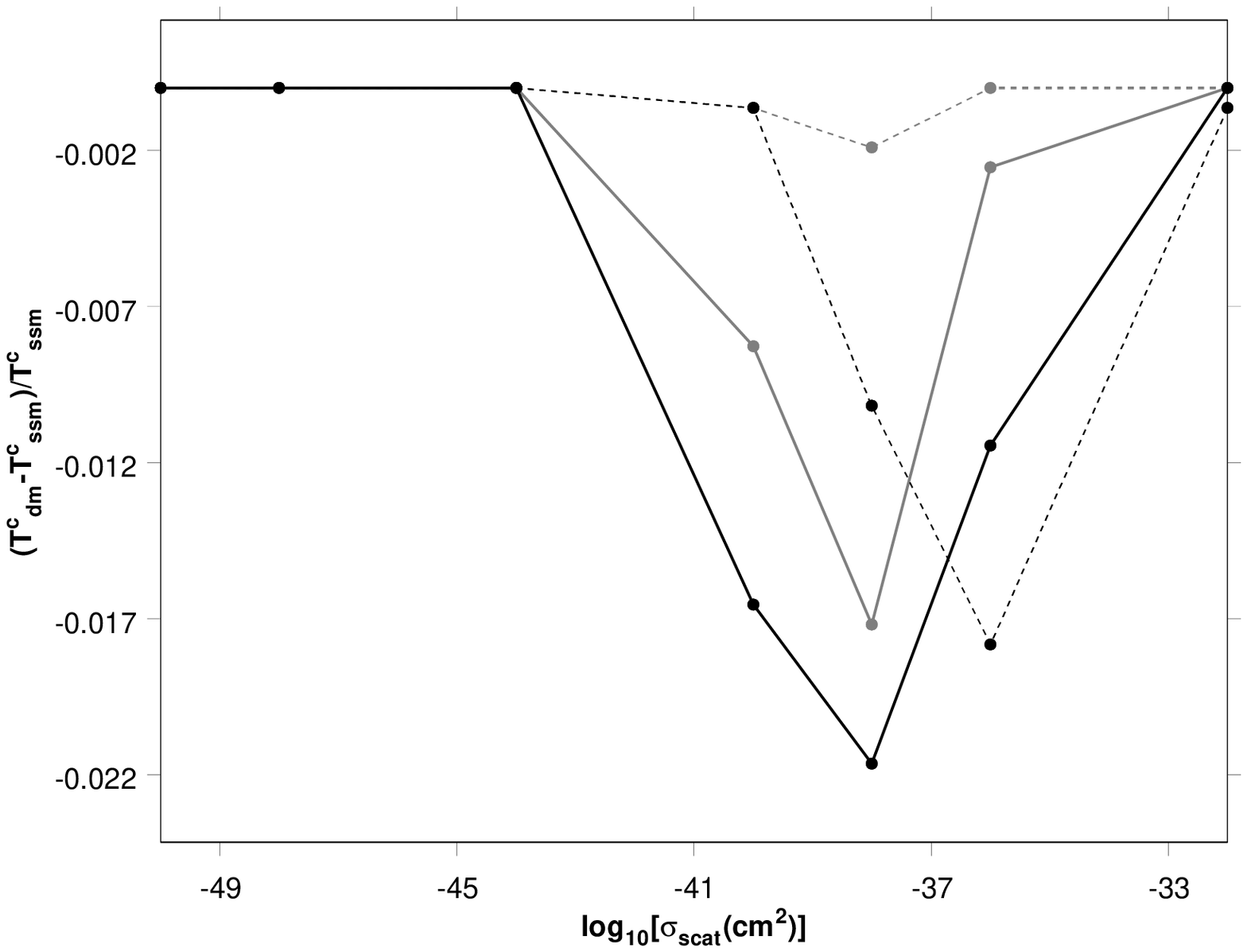,width=10cm,height=7.0cm}}
\vspace{-1.2cm}
\centerline{\psfig{file=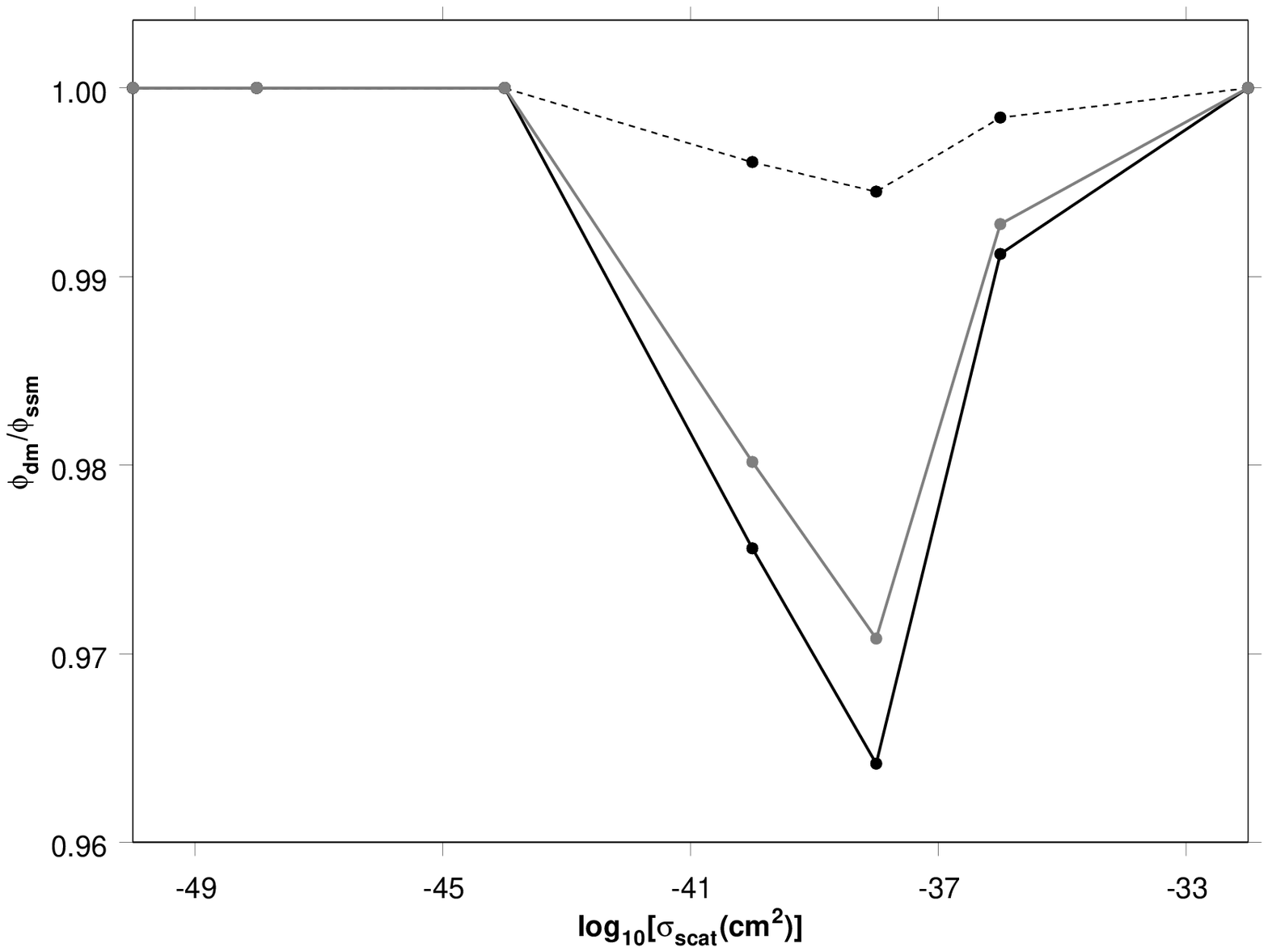,width=10cm,height=7.0cm}}
\caption{\small {\bf a)} The relative differences of the central
temperature between the solar models evolving within a halo of
WIMPs, and the central temperature of the standard solar model.
The different curves correspond to solar models with the following
annihilation cross-sections: $10^{-28}cm^3/s$ (grey dashed curve),
$10^{-32} cm^3/sec$ (grey continuous curve), $10^{-34}cm^3/sec$
(black continuous curve) and $10^{-36}\;cm^3/sec$ (black dashed
curve).{\bf b)} The relative variation on the neutrino flux
predictions to be measured on Earth by the different solar
experiments. The neutrino fluxes prediction of WIMP-accreting
models are normalized to the neutrino fluxes predicted by the
standard solar model.  The different curves correspond to the
variation of the solar neutrino flux predictions for the different
types of neutrino experiments: $^8B$(black continuous curve), $Cl$
(grey continuous curve) and $Ga$ (grey dashed curve).The
WIMP-accreting solar models have been produced by evolving the
star in the presence of an halo of WIMPs of mass of 100 GeV,
annihilation cross-section of $10^{-34}cm^3/s$, and scalar
scattering cross-section between $10^{-32} cm^2$ and
$10^{-50}cm^2$.}
\end{figure}

\begin{figure}[t]
\centerline{\psfig{file=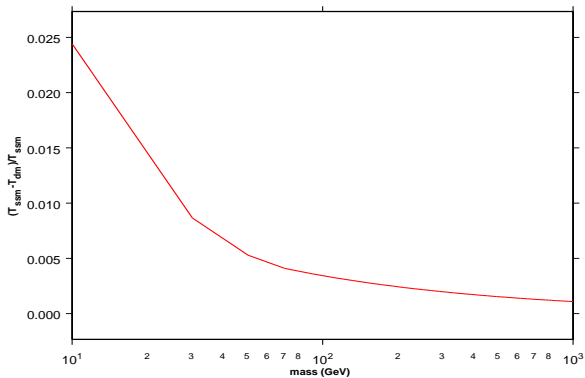,width=9cm,height=6.0cm}}
\caption{\small The relative differences between the central
temperature of the standard solar model and the solar models
within a halo of WIMPs.}
\end{figure}

For comparison, we also estimated the temperature at $10\%$ of the
solar radius, which defines the region within which the inversion
of the radial distribution of the sound speed is not reliable.
Indeed, the nuclear-reacting core is the region of the Sun which
is the most difficult to probe with acoustic modes, because it has
the least influence on the oscillation frequencies (Lopes 2001,
Dziembowski 1996, Gough et al. 1996). It is here that the sound
speed has the highest values and the acoustic wavelengths are the
longest. Although we have access to about 120 modes to explore the
nuclear region, this is quite insufficient to obtain the spatial
resolution needed to successfully invert the sound speed in the
deep solar core. Furthermore, these modes are strongly influenced
by the turbulent motions  in the convection upper layers, as well
as by the surface perturbations of the magnetic field (Lopes \&
Gough 2001). The square of the sound speed is proportional to the
ratio of the temperature to the mean molecular weight, and it
follows that $c_s^2\propto T/\mu$. The recent sound speed results
lead to a small difference between the inverted  square of the
sound speed and the one obtained from the solar standard model,
which is always inferior to $0.6\%$ from the surface towards the
center of the star. In particular, at the location of $10\%$ of
the solar radius, we have $\Delta c_s^2/c_s^2\approx -0.002$
(Turck-Chi\`eze {\it et al.} 2001). If we neglect the variation of
molecular weight at this location, the temperature can be
estimated to be of the order of $13.041\;10^6\;K$, or nearly
$20\%$ of the central temperature of the star. This last point
illustrates how the new results of SNO  significantly constrain
the structure in the region within  $10\%$ of the solar radius,
comparable to the seismic results. Nevertheless, we notice that
the predictions of the standard solar model are limited by small
uncertainties, due to the uncertainties of some physical inputs
and to the different treatments of some physical processes
occurring inside the star, leading to marginal differences in the
central values of the temperature.

It follows from the SNO measurements that any evolutionary model
of the Sun that presents a difference from the central temperature
larger than $5\;10^{-3}$, very likely is not a realistic
representation of the observed Sun. In Fig. 3, we present the
temperature of an isothermal core of WIMPs inside the Sun as a
function of the WIMP mass, for the case where the WIMP
annihilation rate is relatively small and the
annihilation cross-section
is smaller than $10^{-38}cm^3/sec$. It follows that models with
WIMP masses smaller than 60 GeV produce  a variation of the central
temperature relative to the solar standard model that is larger
than the difference presently estimated between the $^8B$ neutrino
flux and the standard solar model. This type of scenario will be
difficult to accommodate in the present context of solar physics.
Nevertheless, the WIMP annihilations will modify their total
number, leading to an increase of the central temperature
and making
it  closer to the temperature predicted by the standard solar
model. This remains true even if the radius of the WIMP core is
nearly independent of the annihilation cross-section.

The other neutrino experiments based on chlorine and gallium,
present alternative methods for inferring the central solar
temperature. In the solar core, the production of $pp$ neutrinos
extends from $1\%$ to $30\%$ of the solar radius,closely following
the production of energy. The $^7Be$ neutrinos are produced in the
region between $3\%$ and $10\%$ of the solar radius,  the maximum
production being at around $5\%$ of the solar radius. To a first
approximation, the production of neutrinos is almost independent
of the solar standard models, and only reflects the nuclear
behaviour of the different nuclear reactions where the neutrinos
are produced. However, some  feedback always occurs due to the
adjustment of the luminosity of the solar models to the observed
luminosity of the present Sun.
 Following Bahcall and Ulrich (1988), we obtain that
 $\phi(pp)\propto T_c^{-1.2}$ and
$\phi(^7Be)\propto T_c^8$. The neutrino flux measured by the
different neutrino detectors probes in different ways  the neutrinos
produced in the nuclear reactions of the pp chain (and CNO cycle).
Indeed, the gallium experiment measured $53\%$ of pp neutrinos,
$26\%$ of $^7Be$ neutrinos, $11\%$ of $^8B$ neutrinos and the rest
from CNO reactions and the $hep$ reactions. The chlorine experiment
measured $78\%$ of $^8B$ neutrinos and $15\%$ of $^7B$ neutrinos.
The different sensitivities of neutrino flux predictions to the
solar model with WIMPs is presented in Fig. 2b. As an
illustrative example for WIMPs with  a mass of 100 GeV, annihilation
cross-section of $10^{-34}\;cm^3/sec$ and scattering cross-section
of $10^{-36}\;cm^2$, the variations in the neutrino fluxes predicted
for the different neutrino experiments present a specific
signature of the isothermal structure of the core. Even if some
physical processes are not correctly implemented or not considered
in the treatment of the solar core, such as the abundance of
chemical elements, the treatment of the screening of electrons in
the nuclear reactions, the treatment of opacity, as well as some
dynamical processes related with the magnetic field and the
transport of energy by gravity waves, in case an isothermal core
is detected by measuring the correct number of neutrinos in the
three types of neutrino experiments, this detection will still
constitute a very strong indication of the existence of WIMPs in
the center of the Sun.


In summary, the results presented here are a first attempt to
search for an indirect indication of the existence of dark matter
in the solar core, and  yields the first predictions of solar
neutrino fluxes in such scenarios. The SNO measurements define a
lower limit for the WIMP mass to be of the order of 60 GeV, in the
case that the annihilating cross section is as small as
$10^{-38}cm^3/sec$. This limit is well above the critical mass of
10 GeV, below which the isothermal core cannot form due to the
fact that WIMPs can escape the Sun's gravitational field.
Furthermore, the quasi-isothermal core of the Sun will create a
peculiar distribution in the solar flux of neutrinos. These
neutrino fluxes from the core can be measured on Earth by the
different solar neutrino experiments. In the case of generic WIMP
masses of order 100 GeV and annihilation cross-sections of order
$10^{-34}\;cm^3/sec$, we expect a decrease in the neutrino fluxes,
as predicted by the standard solar model, by up to $4\%$ for the
chlorine experiments, by $3\%$ for the heavy water experiments and
by $1\%$ for the gallium experiments.

The evolutionary models presented in this Letter exploit the
possibility of the neutrino fluxes being used to scan the space of
parameters of SUSY dark matter particles. Moreover, improved
measurements with  current and forthcoming neutrino experiments
could give us  important and unique insights into the existence of a
possible isothermal core in the center of the Sun, that if detected would
 very
likely be created by  non-baryonic dark matter particles.

The results obtained here highlight the contribution that solar
neutrino measurements on Earth can give us for the understanding
of the evolution of the Sun within the dark matter halo of the
Milky Way. We have  discussed the possible existence of some
processes occurring in the solar core that are missing
from the standard solar model, which in the near
future will be within the reach of  solar neutrino experiments.
This is particularly true for  processes that  originate via
the presence of WIMPs in the core.  In addition,
the possible
detection  of gravity modes by SOHO seismic experiments, such as
GOLF (Turk-Chi\'eze {\it et al.} 2001),  could ultimately
provide a strong
constraint on  the physics of the solar core, and  constitute  an
alternative probe of the solar center that would complement the neutrino
fluxes.

IPL is grateful for support by a grant from Funda\c c\~ao para a
Ci\^encia e T\'ecnologia.

\end{document}